\documentclass[onecolumn,sort&compress]{elsarticle}
\usepackage{ifpdf}
\usepackage[usenames]{color}
\usepackage{bm}
\usepackage{times}
\usepackage{graphicx}
\usepackage{amssymb}
\usepackage{amsmath}
\usepackage{simplewick}
\usepackage{hyperref}
\usepackage{color}

\journal{Computer Physics Communications}

\begin{document}

\begin{frontmatter}

\title{ RCCPAC: A parallel relativistic coupled-cluster program for 
        closed-shell and one-valence atoms and ions in FORTRAN}

\author[a]{B. K. Mani}
\author[b]{ S. Chattopadhyay}
\author[b]{D. Angom \corref{author}}
\cortext[author] {Corresponding author.\\\textit{E-mail address:} 
        angom@prl.res.in }

\address[a]{Department of Physics, 
            Indian Institute of Technology, 
            Hauz Khas, New Delhi 110016, India
            }
\address[b]{Theoretical Physics Division,
            Physical Research Laboratory,
            Navarangpura, Ahmedabad 380 009, Gujarat, India}

\begin{abstract} 
 We report the development of a parallel FORTRAN code, RCCPAC, to solve the 
relativistic coupled-cluster equations for closed-shell and one-valence
atoms and ions. The parallelization is implemented through the use of message 
passing interface, which is suitable for distributed memory computers. The 
coupled-cluster equations are defined in terms of the reduced matrix elements, 
and solved iteratively using Jacobi method. The ground and excited states 
coupled-cluster wave functions obtained from the code could be used to compute 
different properties of closed-shell and one-valence atom or ion. As an example 
we compute the ground state correlation energy, attachment energies, 
$E$1 reduced matrix elements and hyperfine structure constants.

\end{abstract}


\end{frontmatter}

{\bf PROGRAM SUMMARY}

\noindent
{\em Program Title:} RCCPAC                                   \\
{\em Journal Reference:}                                      \\
{\em Catalogue identifier:}                                   \\
{\em Licensing provisions:} none                                   \\
{\em Programming language:}FORTRAN 90                         \\
{\em Computer:} Intel Xeon,                                   \\
{\em Operating system:} General                               \\
{\em RAM:} at least 1.5Gbytes per core.                        \\
{\em Number of processors used:} 4 or higher                              \\
{\em Supplementary material:} none                                 \\
{\em Classification:}                                         \\
{\em External routines/libraries:} none                            \\
{\em Subprograms used:}                                       \\
{\em Journal reference of previous version:}*                  \\
{\em Nature of problem:} Compute the ground and excited state wave functions, 
     correlation energy, attachment energies, and E1 transition amplitude
     and hyperfine structure constant of closed-shell and one-valence atoms or 
     ions using relativistic coupled-cluster theory.  \\
{\em Solution method:} The basic input data required is an orbital basis set 
     generated using the Dirac-Coulomb Hamiltonian. For the present case, 
     closed-shell and one-valence systems, the orbitals are grouped into 
     occupied, valence and virtual. The relativistic coupled-cluster equations 
     of the single and double excitation cluster amplitudes, which form a set of 
     coupled nonlinear equations, are defined in terms of reduced matrix 
     elements of the residual Coulomb interaction. The equations are solved 
     iteratively in parallel using Message Passing Interface (MPI) with the 
     Jacobi method. However, to overcome the slow convergence of the method, we 
     use Direct Inversion in the Iterated Sub-space (DIIS) to accelerate the 
     convergence. For enhanced performance, the two-electron integrals and $6j$-
     symbols are precalculated and stored. Further more, to optimized memory 
     requirements, selected groups of integrals are computed and stored only by 
     the thread which needs the integrals during computation.
\\
{\em Restrictions:} For efficient computations, the two-electron reduced 
     Coulomb matrix elements consisting of three and four virtual states are 
     stored in RAM. This limits the size of the basis set, as there should 
     be sufficient space in RAM to store all the integrals.
   \\
{\em Unusual features:} To avoid replication of data across the cores, and
     optimize the RAM use, the three particle and four particle two-electron
     Coulomb integrals are distributed. That is, following the loop structure 
     in the driver, each core stores only the integrals it requires. With this
     feature there is enormous reduction in the RAM required to store the 
     integrals.
   \\
{\em Additional comments:} The code can be modified, with minimal changes, to
     compute properties other than electromagnetic transitions and hyperfine
     constants with appropriate modifications. The required modfications are
     addition of subroutine to compute the single-electron matrix element, and
     inclusion of calling sequence in the main driver subroutine. 
   \\
{\em Running time: 14 minutes on six processors for the sample case. For
     heavy atoms or ions it could take several days or weeks of CPU time.}\\
   \\


\section{Introduction}

  The coupled-cluster theory (CCT), first developed for applications in 
nuclear physics \cite{coester-58,coester-60}, is one of the most powerful 
quantum many-body theories. The theory was then extended to atomic and 
molecular systems through later developments by \v{C}\'{i}\v{z}ek
\cite{cizek-66,cizek-69} . The theory, as applied to electronic systems, 
is non-perturbative in nature or incorporates electron correlation effects 
to all orders of the electron-electron interactions. In recent years it
has been used with great success in the structure and properties computations
of nuclear \cite{hagen-14}, atomic \cite{geetha-01,pal-07,mani-09}, 
molecular \cite{isaev-04} and condensed matter \cite{li-14} systems. The 
articles in a recent collected volume edited by \v{C}\'{a}rsky, Paldus and 
Pittner \cite{carsky-10} provide very good introduction, and exhaustive 
survey of the recent developments related to the application of CCT in 
various quantum many-body systems. Another reference which elaborates on 
different variants of CCT is the recent review by Bartlett 
and Musia\l{} \cite{bartlett-07}. The canonical version of CCT includes
cluster operators of all possible excitations up to to the number of particles
in the system, however, a truncated scheme which encapsulates all the key
correlation effects is the coupled-cluster singles and doubles (CCSD)
\cite{purvis-82} approximation. This, as the name indicates, includes only
the single and double excitations, but extensive studies have proved the
reliability of the method.

  In the present work we report the development of a computer code which 
implements the relativistic CCT (RCCT) for structure and properties 
computations of closed-shell and one-valence atoms and ions. It is a 
relativistic implementation using the Dirac-Coulomb Hamiltonian, and it must be 
mentioned here that other groups have also used similar implementations for high 
precision structure and properties computations of atoms and ions. These 
include intrinsic electric dipole moments of atoms
\cite{nataraj-08,latha-09}, and parity nonconservation \cite{wansbeek-08},  
hyperfine structure constants \cite{pal-07,sahoo-09} and electromagnetic 
transition properties of atoms and ions. In a series of works
\cite{mani-09,chattopadhyay-12a,chattopadhyay-12b,%
chattopadhyay-13a,chattopadhyay-13b,chattopadhyay-14}, we have 
reported the various approaches adopted to verify the results from the 
present version, and  proof of concept versions of the computer code.
The computer code reported in the present work solves the RCCT equations for 
closed-shell and one-valence atoms and ions using MPI-parallelized Jacobi 
method, and computes the electron correlation energy, attachement energies, 
$E$1 reduced matrix elements, and hyperfine structure constants. 

  An important feature of the code, which optimizes the computational
requirements of the code, is that the loops are structured to achieve minimal
use of the electron-electron integrals, and CCT equations are solved
in cluster driven mode. The basic advantage of such a structure is the 
relative ease of using the code for computations of closed-shell systems,
and one-valence systems. This can be done only through changes in the outer
most few loops. A similar modification with additional computations to 
diagonalize the effective Hamiltonian can be use to adapt the code for 
two-valence sytems. It must, however, be emphasized that the case of 
two-valence systems involves subtle issues related to the model space, and 
require due conceptual considerations, and these are explained in one of our 
previous works \cite{mani-11}.

  A short description of the RCCT is provided in the next section, Section
\ref{rcc_theory}. The section provides a brief, but self contained summary of
the CCSD and linearized RCCT. These are followed by a basic appraisal on how
to compute correlation energy of closed-shell systems using the CCSD wave 
function. The following section describes the Fock-space CCT for one-valence
systems, and how to compute the hyperfine constants and electric dipole 
transition amplitudes using the CC wavefunctions. The next section,
Section \ref{comp_detail} provides crucial information about the grid structure
and type of orbitals used. The Section \ref{detail_impl} contains important 
information on the schemes we have adopted in the code. Some of the 
concepts related to the algorithm adopted add unique features to the present
code, and accounts for optimal usage of memory and computations. One of the key
concepts is  the novel abstraction employed is the structure of the outer 
loops in the implementations. As can seen from the driver subroutines, the
modification from closed-shell to one-valence cluster amplitude computations
involves changes in the driver subroutine only.


\section{RCC theory of closed-shell systems}
\label{rcc_theory}

    The Dirac-Coulomb Hamiltonian $ H^{\rm DC}$ is an appropriate Hamiltonian 
to account for the relativistic effects in the structure and properties
calculations of atoms and ions. For an $N$ electron atom or ion, in atomic
units ($\hbar=m_e=e=4\pi\epsilon_0=1$),
\begin{equation}
  H^{\rm DC}=\sum_{i=1}^N\left [c\bm{\alpha}_i\cdot \mathbf{p}_i+
             (\beta_i-1)c^2 - V_N(r_i)\right ] +\sum_{i<j}\frac{1}{r_{ij}},
  \label{dchamil}
\end{equation}
where $\bm{\alpha}_i$ and $\beta$ are the Dirac matrices, $\mathbf{p}$ is the
linear momentum, $V_N(r)$ is the nuclear Coulomb potential, and the last term
is the electron-electron Coulomb interactions. For a closed-shell system, the
ground state satisfies the eigenvalue equation
\begin{equation}
  H^{\rm DC} |\Psi_0\rangle = E_0 |\Psi_0\rangle,
\end{equation}
where $|\Psi_0\rangle$  and $E_0$ are the ground state exact wave function 
and  energy, respectively. In the RCC theory, the exact ground state wave 
function
\begin{equation}                                                             
  |\Psi_0\rangle = e^T|\Phi_0\rangle,                                
  \label{cc_close}                                                             
\end{equation}                                                               
where $T$ is the CC operator for closed-shell systems and $|\Phi_0\rangle$ is
the Dirac-Hartree-Fock reference state. For the present case, the CC 
operator is
\begin{equation}
  T= \sum_{i=1}^{N} T_i,
  \label{t_def}
\end{equation}
here, the index $i$ indicates the level of excitation. The eigenvalue equation 
using the normal form of an operator 
$O_N = O - \langle\Phi_0|O|\Phi_0\rangle$ can be rewritten as
\begin{equation}                                                             
  H_Ne^T|\Phi_0\rangle  = \Delta E e^T|\Phi_0\rangle,
  \label{ccsd_eq}
\end{equation}
where $H_N = H^{\rm DC} - \langle\Phi_0|H^{\rm DC}|\Phi_0\rangle$ is the 
normal form of $H^{\rm DC}$ and 
$\Delta E = E_0 - \langle\Phi_0|H^{\rm DC}|\Phi_0\rangle$ is the correlation 
energy. Multiplying the equation from left by $e^{-T} $, and projecting on
$\langle\Phi_0|$,  and the excited states $\langle\Phi^*| $, we get 
\begin{subequations}
  \label{cc_sing-doub}
\begin{eqnarray}
  \langle\Phi_0|e^{-T}H_{\rm N}e^T|\Phi_0\rangle &= &\Delta E,
  \label{cc_corr}    \\
  \langle\Phi^*|e^{-T}H_{\rm N}e^T|\Phi_0\rangle &=& 0.
  \label{cc_amp}
\end{eqnarray}
\end{subequations} 
The first equation gives the correlation energy, and the second 
equation is a set of coupled nonlinear equations for the cluster amplitudes. For compact
notation, define $\bar{H}_{\rm N}  =  e^{-T}H_{\rm N}e^T$  as the dressed or 
the similarity transformed Hamiltonian. As $H^{\rm DC}$ consists of only
one- and two-body terms, and following Wick's theorem we can write
\begin{equation}
  \bar{H}_{\rm N}=H_{\rm N}
    + \bigg \{\contraction[0.4ex]{}{H}{_{\rm N}}{T} H_{\rm N}T\bigg \} +
    + \frac{1}{2!}\bigg \{\contraction[0.4ex]{}{H}{_{\rm N}}{T}
      \contraction[0.7ex]{}{H}{_{\rm N}T}{T} H_{\rm N}TT\bigg \} 
    + \frac{1}{3!}\bigg \{\contraction[0.4ex]{}{H}{_{\rm N}}{T}
      \contraction[0.7ex]{}{H}{_{\rm N}T}{T}
      \contraction[1.0ex]{}{H}{_{\rm N}TT}{T}
      H_{\rm N}TTT\bigg \}  
    + \frac{1}{4!}\bigg \{\contraction[0.4ex]{}{H}{_{\rm N}}{T}
      \contraction[0.7ex]{}{H}{_{\rm N}T}{T}
      \contraction[1.0ex]{}{H}{_{\rm N}TT}{T}
      \contraction[1.3ex]{}{H}{_{\rm N}TTT}{T}
      H_{\rm N}TTTT\bigg \},
  \label{hn_bar}
\end{equation}
where $\contraction[0.4ex]{}{A}{\cdots}{B}A\cdots B$ represents contraction
between the operators $A$ and $B$, and $\{\cdots\}$ denote the operators
are normal ordered. This shows, despite the exponential nature of the ansatz, 
the dressed Hamiltonian consists of terms up to quartic order in $T$, and 
hence, the CCT equations contains terms of fourth order as the highest 
degree of non-linearity.
%
%
\begin{center}
\begin{figure}[h]                                                               
  \centering
  \includegraphics[width = 9.0 cm]{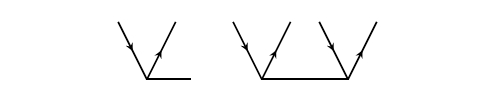}                                 

  \caption{Diagrammatic representation of the closed-shell CC operators
           $T_1$ and $T_2$. The orbital lines with up (down) arrows indicate 
           particle (hole) states. The horizontal solid line represents the
           electron-electron Coulomb interaction to all orders.}
  \label{t1t2_fig}
\end{figure}                                                                 
\end{center}                                                               
%
%


\subsection{Coupled-cluster singles and doubles approximation}

  In CCT, the number of cluster amplitudes increase exponentially with 
order of excitation $i$ in Eq. (\ref{t_def}). So, for systems with large $N$, 
it is nontrivial to include all the cluster amplitudes of all possible 
excitations. An approximation which encapsulates a major part of the 
correlation effects is the CCSD approximation \cite{purvis-82}, in which
we retain only $T_1$ and $T_2$. Going beyond CCSD by inclusion of $i>2 $ is 
nontrivial and computationally resource intensive. For closed-shell systems, 
the CCSD gives an accurate description of the structure and properties.
Using the occupation number representation and in normal ordered form, the
cluster operators are
\begin{subequations}
\begin{eqnarray}
  T_1 & = &\sum_{a, p}t_a^p a_p^{\dagger}a_a, \\                        
  T_2 & = &\frac{1}{2!}\sum_{a, b, p, q}t_{ab}^{pq}
           a_p^{\dagger}a_q^{\dagger}a_ba_a,
\end{eqnarray}
\end{subequations}
where $t_{\cdots}^{\cdots}$ represent the closed-shell CC amplitudes, and the 
indices $abc\ldots (pqr\ldots)$ represent the core (excited) single particle
states. Replacing $\langle\Phi^*|$ in Eq. (\ref{cc_amp}) with 
$\langle\Phi^p_a|$ and $\langle\Phi^{pq}_{ab}|$, which are the single and 
double excited unperturbed or Dirac-Hartree-Fock states, respectively, we 
get the cluster equations for $T_1$ and $T_2$. So, the CC amplitudes
$t_a^p$ and $t_{ab}^{pq}$ are solutions of the coupled nonlinear equations
\begin{subequations}
\begin{eqnarray}
  \langle\Phi^p_a|\bar{H}_{\rm N}|\Phi_0\rangle = 0,
  \label{cc_sing}    \\
  \langle\Phi^{pq}_{ab}|\bar{H}_{\rm N}|\Phi_0\rangle = 0.
  \label{cc_doub}
\end{eqnarray}
\end{subequations} 
The general details of the derivations, in the context of non-relativistic
description, are given in ref. \cite{lindgren-86,shavitt-09}. For RCC, the 
relevant details for closed-shell systems
is described in our previous work \cite{mani-09}. To simplify the evaluation 
of the terms in the CC equations we use Goldstone diagrams, and angular
integration are performed using angular momentum diagrams. The diagrammatic
representation of cluster operators $T_1$ and $T_2$ are as shown in 
Fig. \ref{t1t2_fig}. In the present work, for the diagrammatic analysis, we 
follow the conventions and notations in ref. \cite{lindgren-86}.
%
%
\begin{figure}[h]
  \centering
  \includegraphics[width = 9.0 cm]{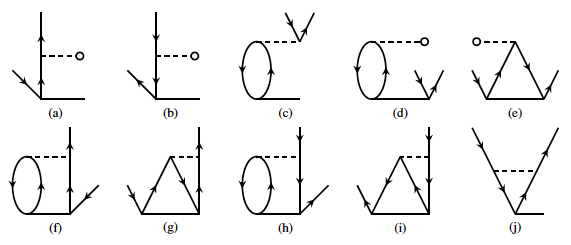}                                 
  \caption{Diagrams which contribute to the linearized RCC for singles 
           ($T_1$). Dashed lines represent the residual Coulomb interaction,
           and the solid lines are the cluster operators.}                  
 \label{lcc_fig_sing}
\end{figure}
%
%


\subsection{Linearized RCC}

\begin{figure}[h]
  \centering
  \includegraphics[width = 9.0 cm]{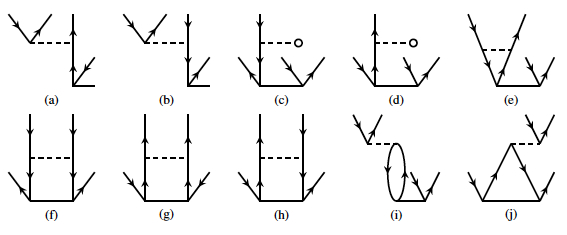}                                 
  \caption{Diagrams which contribute to the linearized RCC for doubles
           ($T_2$). Dashed lines represent the residual Coulomb interaction,
           and the solid lines are the cluster operators.}                  
 \label{lcc_fig_doub}
\end{figure}

    The dressed Hamiltonian $\bar{H}_{\rm N}$, as given in 
Eq. (\ref{hn_bar}) have contributions from different orders of $T$, up to
quartic. Hence, the Eqs. (\ref{cc_sing}) and (\ref{cc_doub}) form a
set of  non-linear coupled algebraic equations. However, an approximation 
often used as a starting point of RCC computations is the linearized RCC, 
where only the linear terms are retained in the computations.  In this 
approximation, the dressed Hamiltonian is
\begin{equation}                   
  \bar{H}_{\rm N} = H_{\rm N} 
     + \bigg \{\contraction[0.4ex]{}{H}{_{\rm N}}{T}H_{\rm N} T\bigg \}.
  \label{hbar_lin}
\end{equation}                                                               
Using Eq. (\ref{hbar_lin}), from Eq. (\ref{cc_sing-doub}) single and double 
RCC equations are then
\begin{subequations}
\begin{eqnarray}
  \langle\Phi^p_a|H_{\rm N} 
  + \bigg \{\contraction[0.4ex]{}{H}{_{\rm N}}{T}H_{\rm N} T\bigg \}
    |\Phi_0\rangle = 0, \\
  \langle\Phi^{pq}_{ab}|H_{\rm N}
   + \bigg \{\contraction[0.4ex]{}{H}{_{\rm N}}{T}H_{\rm N}T\bigg \}
   |\Phi_0\rangle = 0.                                                 
\end{eqnarray} 
\end{subequations} 
For the CCSD approximation $T = T_1 + T_2$, these equations are then
\begin{subequations}
 \label{lcc_sing-doub}
\begin{eqnarray}
  \langle\Phi^p_a|\bigg \{\contraction[0.4ex]{}{H}{_{\rm N}}{T}H_{\rm N} T_1
    \bigg \} +  \bigg \{\contraction[0.4ex]{}{H}{_{\rm N}}{T}H_{\rm N} T_2
    \bigg \}|\Phi_0\rangle &=&  
  -\langle\Phi^p_a|H_{\rm N}|\Phi_0\rangle, \\
   \langle\Phi^{pq}_{ab}|\bigg \{\contraction[0.4ex]{}{H}{_{\rm N}}{T}H_{\rm N}
   T_1\bigg \} +    \bigg \{\contraction[0.4ex]{}{H}{_{\rm N}}{T}H_{\rm N} 
   T_2\bigg \}|\Phi_0\rangle &=&
  -\langle\Phi^{pq}_{ab}|H_{\rm N} |\Phi_0\rangle.
\end{eqnarray}
\end{subequations}
The CC diagrams which contribute to $T_1$ and $T_2$ in the above equations
are shown in the Figs. \ref{lcc_fig_sing} and \ref{lcc_fig_doub}, 
respectively. Up to this 
point we have used the notation
$\contraction[0.4ex]{}{A}{\ldots}{B}A\ldots B$ to represent contraction 
between two operators $A$ and $B$. Here after we drop this explicit
notation, and the contractions are implied in expressions
with products of operators.
%

The other form of RCC equations is to write in terms of cluster amplitudes, 
$t^p_a$ and $t^{pq}_{ab}$. The linearized RCC equation of $T_1$  is then
\begin{equation}
  \epsilon_a^p t^p_a = \sum_{bq}\tilde{v}^{bp}_{qa} t^q_b +
  \sum_{bqr}\tilde{v}^{bp}_{qr} t^{qr}_{ba}
  -\sum_{bcq} v^{bc}_{qa} \tilde{t}^{qp}_{bc},
  \label{lin_sing}
\end{equation}
where $\epsilon_a^p = \epsilon_a - \epsilon_p $, $v^{ij}_{kl}$ is the 
matrix element of electron-electron Coulomb interaction
$\langle ij|(1/r_{12})|kl\rangle$, and 
$\tilde{v}^{ij}_{kl}=v^{ij}_{kl} - v^{ij}_{lk} =v^{ij}_{kl} - v^{ji}_{kl}$ 
is the antisymmetrized matrix element. Similarly, the compact notation the 
antisymmetrized CC amplitude is ${\widetilde t}^{ij}_{kl}$. Like the $T_1$, 
the linearized RCC equation for $T_2$ is
\begin{eqnarray}
  \epsilon_{ab}^{pq} t^{pq}_{ab} &=& v^{pq}_{ab} + \bigg [ 
       \sum_{r} v^{pq}_{rb} t^r_a  - \sum_{c} v^{cq}_{ab} t^p_c
       +\sum_{rc} \Big ( v^{pc}_{ar} \tilde {t}^{rq}_{cb}
    - v^{pc}_{rb} t^{rq}_{ac} - v^{cp}_{ar} t^{rq}_{cb}  
       \Big ) \bigg ] 
                          \nonumber \\
  &&+\left [ \begin{array}{c}
           p\leftrightarrow q \\
           a\leftrightarrow b
           \end{array} \right ] 
    +\sum_{rs} v^{pq}_{rs} t^{rs}_{ab}
    +\sum_{cd} v^{cd}_{ab} t^{pq}_{cd},
  \label{lin_doub}
\end{eqnarray}
where 
$ \epsilon_{ab}^{pq} = \epsilon_a + \epsilon_b - \epsilon_p - \epsilon_q$ and
$\bigl[ \begin{smallmatrix}p\leftrightarrow q \\ a\leftrightarrow b
\end{smallmatrix} \bigr ]$ represents terms similar to those within
parenthesis but with the combined permutations $p\leftrightarrow q$ and
$a\leftrightarrow b$. These equations are the algebraic equivalent of the 
diagrams shown in Figs. \ref{lcc_fig_sing} and \ref{lcc_fig_doub}, respectively.
The evaluations are based on the rules to analyze Goldstone diagrams. It is 
the preferred scheme as the angular momentum diagram evaluation is easier. 
At the implementation level, the angular integrals are evaluated so that
the equations are in terms of reduced matrix elements of $T$ operators. This
minimizes the number of cluster amplitudes and simplify the computations.
%
%
\begin{figure}[h]
  \centering
  \includegraphics[width = 9.0 cm]{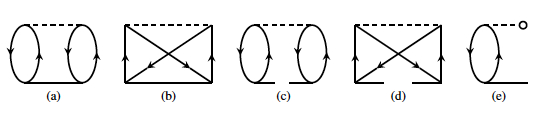}                                 
  \caption{The coupled-cluster diagrams which contribute the correlation 
           energy for closed-shell systems. The dashed lines represent the
           Coulomb interaction. The solid lines, however, are to represent the
           CC operators.}                   
 \label{delta_e_fig}
\end{figure}

%
%


\subsection{Correlation energy}

    The correlation energy of a closed-shell system, as defined in 
Eq. (\ref{cc_corr}), is the expectation value of $\bar{H}_N$ with respect 
to $|\Phi_0\rangle$. That is     
\begin{equation}                                                               
  \Delta E = \langle\Phi_0|\bar{H}_{\rm N}|\Phi_0\rangle,
 \label{delta_e}
\end{equation}                                                                 
and in the CCSD approximation it has contributions from $T_1$ and $T_2$. The
diagrams which contribute to $\Delta E$ are shown in the 
Fig. \ref{delta_e_fig}. The dominant contributions are from the
diagrams (a) and (b), which is natural as the $t_{ab}^{pq}$  are larger in 
value than $t_a^p$. The other two diagrams, (c) and (d), arise from terms 
which are second order in $T_1$ and have smaller contribution.  The last 
diagram Fig. \ref{delta_e_fig}(e), following Koopman's theorem, is
zero when Dirac-Hartree-Fock orbitals are used. Neglecting this diagram, the 
algebraic expression of $\Delta E$ corresponding to the first four diagrams in 
Fig. \ref{delta_e_fig} is
\begin{equation}
  \Delta E = \tilde{v}_{pq}^{ab}\Big ( t_{ab}^{pq} + t_a^pt_b^q \Big ).
\end{equation}
This can be computed once the cluster amplitudes are known. Albeit, the 
correlation equation is written first in eq. \ref{cc_sing-doub}, but in 
computations, it is evaluated later.


\section{Fock-space CC theory and properties of one-valence systems}
\label{rcc_1v}

\begin{figure}[h]
  \centering
  \includegraphics[width = 9.0 cm]{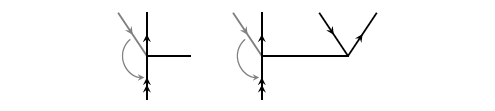}                                 
  \caption{Diagrammatic representation of the $S_1$ and $S_2$ cluster 
           operators. These can be considered as topological transformations
           of $T_1$ and $T_2$ cluster operators with one of the core orbital
           lines rotated downwards as indicated by the arrow and orbital line
           in gray.
           }
  \label{s1s2_fig}
\end{figure}

  The key difference of one-valence atom or ions from the closed-shell ones 
is the presence of a single electron in the outer most or the valence shell. 
To account for the correlation effects arising from the valence electron, 
we use Fock-space coupled-cluster theory and introduce a new set of cluster 
operators $S$. In the CCSD approximation $S =S_1 + S_2 $ and these are 
defined as 
\begin{subequations}
\begin{eqnarray}
  S_1 & = &\sum_{p}s_v^p a_p^{\dagger}a_v, \\                        
  S_2 & = &\frac{1}{2!}\sum_{a, p, q}s_{va}^{pq}
           a_p^{\dagger}a_q^{\dagger}a_aa_v,
\end{eqnarray}
\end{subequations}
where, $v$ is the index which identifies the valence electron and 
$s_{\ldots}^{\ldots}$ are the cluster amplitudes corresponding to the 
valence sector. In the Fock-space coupled-cluster of one-valence systems, 
as the name indicates, the starting point of the calculation is the 
closed-shell coupled-cluster. Based on the Hilbert space of the closed-shell 
system, we generate the one-valence Hilbert space by adding an electron. 
These two Hilbert spaces together form the Fock-space for the one-valence
system. Thus, the reference state of the one-valence system, starting from 
the closed-shell system, is $|\Phi_v\rangle = a_v^{\dagger}|\Phi_0\rangle$. 
Here, recall that $|\Phi_0\rangle$ is the Dirac-Hartree-Fock state of the 
closed-shell system and $a_v^{\dagger}$ adds the valence electron. The exact 
state of the system is 
\begin{equation}
  |\Psi_{v}\rangle = e^{(T + S)}|\Phi_{v}\rangle  = e^T(1+S)|\Phi_{v}\rangle .
  \label{onev_exact}
\end{equation}
It is to be noted that the closed-shell part, which involves $T$, are 
calculated to all orders, or we retain the exponential form of the CCT in 
the closed-shell sector, but $S$ restricted to linear terms only.  This is 
due to the presence of a single valence electron, and the diagrammatic 
representations of $S$  are shown in Fig. \ref{s1s2_fig}.
The schematic representation of converting the computation of $T$ to $S$ 
cluster amplitudes is shown in the figure. It is equivalent to converting
one of the core orbitals in the driver programs to valence orbital. Once the
cluster amplitudes are obtained, the coupled-cluster wavefunctions can be 
used for properties computations. For atoms or ions there are, in general,
two classes of properties. First, the properties associated with a state
which are calculated as expectations, and second, transition properties
associated with an initial and final states. The hyperfine structure
constants, and electric dipole $E1$ transition properties are described as
examples of the former and latter classes, respectively.


\subsection{One-valence coupled-cluster equations}

The one-valence exact state $|\Psi_v\rangle$ satisfies the Schr\"{o}dinger 
equation 
\begin{equation}
   H^{\rm DC} e^{T}(1+S)|\Phi_v\rangle = E_v |\Phi_v\rangle,
\end{equation}
where $H^{\rm DC}$ is now the Dirac-Coulomb Hamiltonian in the one-valence
sector. Projecting this equation on $e^T$, and using the normal form of the
Hamiltonian the one-valence cluster amplitudes, in the CCSD approximation, 
are solutions of the coupled linear equations
\begin{subequations}
\begin{eqnarray}
  \langle \Phi_v^p|\bar H_N \! +\! \{\contraction[0.5ex]
  {\bar}{H}{_N}{S} \bar H_N S\} |\Phi_v\rangle
  &=&E_v^{\rm att}\langle\Phi_v^p|S_1|\Phi_v\rangle ,
  \label{ccsingles}     \\
  \langle \Phi_{va}^{pq}|\bar H_N +\{\contraction[0.5ex]
  {\bar}{H}{_N}{S}\bar H_N S\} |\Phi_v\rangle
  &=& E_v^{\rm att}\langle\Phi_{va}^{pq}|S_2|\Phi_v\rangle,
  \label{ccdoubles}
\end{eqnarray}
\label{cct_1v}
\end{subequations}
where, $E_v^{\rm att}$ is the attachment energy of the valence shell or the 
energy released when an electron is attached to the open shell $v$ in the 
closed-shell ion. The excited determinants $|\Phi^p_v\rangle$ and 
$|\Phi^{pq}_{va}\rangle$, like in the case of closed-shell system, are
obtained by exciting one and two electron from the reference
state $|\Phi_v\rangle$. By definition
\begin{equation}
   E_v^{\rm att} = E_v - E_0,
\end{equation}
where $E_v = \langle \Phi_v|\bar H_N + \{\contraction[0.5ex]
  {\bar}{H}{_N}{S} \bar H_N S\} |\Phi_v\rangle$  and
$E_0=\langle\Phi_0|\bar H |\Phi_0\rangle$ are the exact energies of 
states $|\Psi_v\rangle$ and $|\Psi_0\rangle$ respectively. A detailed 
description of the derivation and interpretations of these equations
are given ref \cite{Mani-10}. One key difference of Eqs. \ref{cct_1v} from the
closed-shell case are the terms on the right hand side of the equations, and 
these are the renormalization terms. In diagrammatic representation, these are
the folded diagrams, and have very different topological structure from the
$\bar H$ or $\contraction[0.5ex] {\bar}{H}{_N}{S}\bar H_N S$.
\begin{figure}[t]
  \centering
  \includegraphics[width = 9.0 cm]{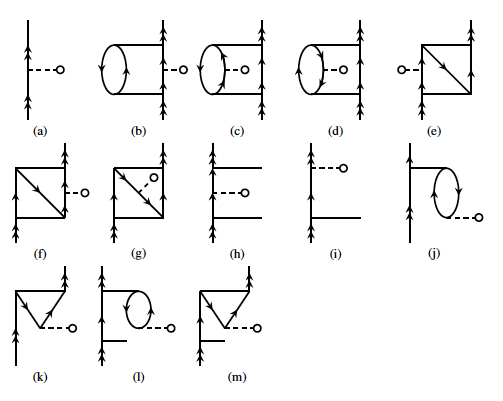}                                 
  \caption{Few of the diagrams with dominant contributions to the hyperfine 
           structure constants in one-valence systems. In the diagrams, the 
           dashed lines marked with open circle represents hyperfine 
           interaction.}
  \label{hfs_diagrams}
\end{figure}


\subsection{Hyperfine Structure Constants of one-valence systems}

 For atom or ions in the state $|\Psi_v\rangle $ the experimentally measured 
property $A$ is the expectation 
\begin{equation}  
  \langle A \rangle = \frac{\langle \Psi_v|A|\Psi_v \rangle}
                      {\langle \Psi_v |\Psi_v \rangle}.
  \label{a_expect}
\end{equation} 
The property could be associated with either an interaction which is internal 
or in response to an external perturbation. In the present case, we consider 
$A$ as the hyperfine interaction $H_{\rm hfs}$ which is internal to the atom or 
ion. It arises from the coupling of the nuclear electromagnetic moments to 
the electromagnetic field of the electrons. The total angular momentum of the
system is then $F = I + J$, where $I$ and $J$ are the nuclear spin, and 
total angular momentum of the electrons. The states of the system are 
represented as $|(IJ)FM_F\rangle$, and the form of the hyperfine 
interaction is \cite{schwartz-55}
\begin{equation}
  H_{\rm hfs} = \sum_i\sum_{k, q}(-1)^q t^k_q(\hat {\bf r}_i) T^k_{-q},
  \label{hfs_ham}
\end{equation}
where $t^k_q(\bm{r})$ and $T^k_{q}$ represent irreducible tensor operators 
of rank $k$ in the electron and nuclear spaces, respectively, and index $i$
is summed over all the electrons in the system. Following the parity 
considerations, only the even and odd values of $k$ are possible for the 
electric and magnetic interactions, respectively. In general, the parameters 
which represent the energy shift due to hyperfine interactions are the 
hyperfine structure constants. For one valence systems, the magnetic dipole 
hyperfine structure constant is
\begin{equation}
  a = \frac{g_I\mu_N}{\sqrt{j_v(j_v+1)(2j_v+1)}}
      \langle \Psi_v ||\sum_it^1(\mathbf{r}_i)||\Psi_v \rangle,
      \label{hfs_mdipole}
\end{equation}
where,  $j_v$ is the total angular momentum of the valence electron, 
$g_I$ is the gyromagnetic ratio and $\mu_N$ is the nuclear magneton. 
In terms of the coupled-cluster wave functions, the reduced matrix element
of the hyperfine interaction Hamiltonian is
\begin{equation}
  \langle \Psi_v\parallel H_{\rm hfs}\parallel \Psi_v \rangle = 
       \langle \Phi_v\parallel \tilde H_{\rm hfs} 
       + 2 S^\dagger \tilde H_{\rm hfs}
       + S^\dagger \tilde H_{\rm hfs} S \parallel \Phi_v\rangle ,
  \label{hfs_num}
\end{equation}
where, $\tilde H_{\rm hfs} = e{^T}^\dagger H_{\rm hfs} e^T$ is the dressed 
operator. We arrive at the factor of two on the second term on the right hand 
side as  $S^\dagger \tilde H_{\rm hfs} = \tilde H_{\rm hfs} S $. A 
convenient form of $\tilde H_{\rm hfs}$ is
\begin{equation}
  \tilde H_{\rm hfs} = H_{\rm hfs} e^T + \sum_{n = 1}^\infty \frac{1}{n!}
                 \left ( T^\dagger \right )^n H_{\rm hfs} e^T,
  \label{A_tilde}
\end{equation}
and the normalization factor is
\begin{equation}
  \langle \Psi_v |\Psi_v \rangle = 
      \langle \Phi_v|\left (1 + S^\dagger\right ) e{^T}^\dagger e^T
                     \left ( 1 +  S\right )|\Phi_v\rangle.
\end{equation}
In the computations we consider the first few terms in order of the
cluster operators from the non-terminating series of $ \tilde H_{\rm hfs}$, 
and the operator $ e{^T}^\dagger e^T$ in the normalization factor. As example,
the diagrams corresponding to dominant terms are shown in 
Fig. \ref{hfs_diagrams}.


\subsection{Electric dipole transition amplitudes for one-valence systems}

 The electromagnetic transition amplitudes is another class of properties
of atoms or ions which involve two states, an initial and final state. Among
the various electromagnetic multipole transitions, the electric dipole $E1$ is 
the most dominant, and occures between two states of opposite parities. 
In terms of theoretical description, the important quantity related to 
$E1$ transition between the initial and final states $|\Psi_i\rangle$ and 
$|\Psi_f\rangle$, respectively, is 
\begin{equation}
  D_{if} = \frac{\langle \Psi_f\parallel \mathbf{D}\parallel \Psi_i\rangle}
           {\sqrt{\langle \Psi_f|\Psi_f\rangle\langle\Psi_i |\Psi_i\rangle}},
  \label{d_if}
\end{equation}
where $\mathbf{D}$ is the electric dipole operator. To simplify the expression
we can partition the coupled-cluster wave operator as
\begin{equation}
  e^{T}(1+S) = \Omega  = \Omega ^+ + \Omega ^-.
\end{equation} 
Where $\Omega ^+ $ and $ \Omega ^-$ are the components of the wave operator 
which operates on the even and odd parity reference states. We can, then,
write
\begin{equation}
  D_{if} = \frac{\langle \Psi_f^0\parallel{\Omega^{\mp}}^\dagger D\Omega^{\pm}
           \parallel\Psi_i^0\rangle}
           {\sqrt{\langle \Psi_f|\Psi_f\rangle\langle\Psi_i |\Psi_i\rangle}}.
\end{equation}
For the one-valence system if $|\Psi_v\rangle$ and $|\Psi_w\rangle$ are the 
initial and final states, respectively, then the reduced matrix element is
\begin{equation}
  \langle \Psi_w\parallel \mathbf{D}\parallel\Psi_v \rangle = 
       \langle \Phi_v\parallel \tilde {\mathbf{D}} 
        + S^\dagger \tilde {\mathbf{D}} + \tilde{\mathbf{ D}} S 
        + S^\dagger \tilde {\mathbf{D}} S \parallel\Phi_v\rangle .
  \label{dip_1v}
\end{equation}
Albeit the expressions are similar to Eq.(\ref{hfs_num}), there is one 
important difference. Unlike in the case of hyperfine structure constant
$S^\dagger \tilde{\mathbf{ D}} \neq \tilde{\mathbf{ D}}S$ as 
the $|\Psi_v\rangle$ and $|\Psi_w\rangle$ are different states. The code in
the present work computes the $E$1 reduced matrix elements as the $E$1
transition properties can be obtain from it in combination with the excitation 
energies. In terms of diagrams, we can obtain the dominant contributions after 
appropriate modification of the diagrams in Fig. \ref{hfs_diagrams}. And, the 
modifications are: changing $H_{\rm hfs}$ to $\mathbf{D}$, and relabelling the 
final state as the valence state $v$.


\section{Computational details}
\label{comp_detail}

\subsection{Radial grid}

   For numerical evaluation of the two-electron Slater integrals, the radial
wave functions are defined in an exponential grid. So that the $i$th radial
grid point has the value, in atomic units,
\begin{equation}
  r(i) = r_0 \left [ e^{(i-1)h} - 1 \right ],
\end{equation}
where, for the present work, we use $r_0=2.0\times 10^{-6}$ and $h=0.05$. This
choice of radial grid representation samples the nuclear Coulomb potential 
very well: smaller separation in the $r\ll1$ and larger separation at 
$r\gg 1$, where the potential is strong and weak, respectively. This choice
is similar to the grid used in GRASP2K \cite{jonsson-07}. In the present
implementation of the code, the details of the grid are read from the
orbital basis file. The obvious advantage of this implementation is the 
consistent choice of grid parameters across codes as we generate the basis
set using another code.


\subsection{Orbital basis set}

 The single particle state $\psi_{n\kappa m}$ with principal quantum 
number $n$, relativistic total quantum number $\kappa$, and magnetic quantum 
number $m$ is defined as the four-component spin-orbital
\begin{equation}
  \psi_{n\kappa m}(\bm{r})=\frac{1}{r}
  \left(\begin{array}{r}
            P_{n\kappa}(r)\chi_{\kappa m}(\mathbf{r}/r)\\
           iQ_{n\kappa}(r)\chi_{-\kappa m}(\mathbf{r}/r)
       \end{array}\right),
\end{equation}
where $P_{n\kappa}(r)$ and $Q_{n\kappa}(r)$ are the large and small component
of the radial wave functions, respectively, and 
$\chi_{\kappa m}(\mathbf{r}/r)$ are the spinor spherical harmonics. In the 
example calculations, the radial functions are even tempered Gaussian type 
orbitals (GTOs) \cite{mohanty-90} on a grid \cite{chaudhuri-99}. The large 
component $P_{n\kappa}(r) $ are then linear combination of the Gaussian type 
functions
\begin{equation}
   g_{\kappa p}^{L}(r) = {\rm N}^{L}_{\kappa p} r^{n_{\kappa}}
                         e^{-\alpha_{p}r^{2}},
\end{equation}
where,  $ {\rm N}^{L}_{\kappa p}$ is the normalization constant, and 
$\alpha_{p}$ is the exponent. The exponents are defined in terms of 
two parameters and forms a geometric series $\alpha_p = \alpha_0 \beta ^{p-1}$,
where $\alpha_0$ and $\beta$ are two constants. The choice of these constants
are optimized to matched the self consistent field and single particle energies
obtained from GRASP2K \cite{jonsson-07}. The large component can then be
written as 
\begin{equation}
   P_{n\kappa m}(r) = \sum_p^{n_\kappa} C^L_{\kappa p} g_{\kappa p}^{L}(r),
\end{equation}
where $ C^L_{\kappa p}$ is the coefficient of linear combination and 
$n_\kappa$ is the number of Gaussian type functions considered for the 
symmetry. Using the GTOs, the Slater integrals or the two-electron Coulomb
interaction matrix elements are computed using the subroutines from GRASP2K
\cite{jonsson-07}.
%
%
\begin{figure}[h]
  \centering
  \includegraphics[width = 9.0 cm]{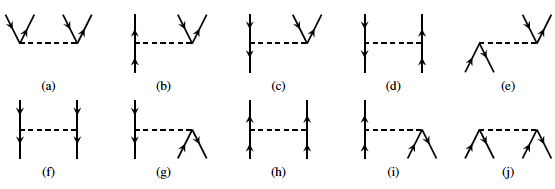}                                 
  \caption{Diagrams of the ten topologically unique representations of the
           Slater integrals. All the diagrams contribute to the $T_2$ equation,
           but the first three diagrams (a-c) do not contribute to $T_1$.
           }                  
 \label{slater_fig}
\end{figure}
%
%


\section{Details of implementation}
\label{detail_impl}

\subsection{Loop structures}

  To aid the conversion from closed-shell to open shell calculations, the 
implementation has a two-tier structure. These translate to heirarchies of 
loop structures in terms of the orbitals. For single excitation cluster 
operator $t_a^p$ , the first tier consists of loops corresponding to the free 
orbital lines $p$ and $a$. Similarly, for double excitation cluster 
operators $t_{ab}^{pq}$, the $p$, $q$, $a$ and $b$ form the outer loops. We 
refer to this outer loop structure as {\em cluster } driven. 

 To describe the second tier of loops, we classify the Slater 
integrals into different categories. In total there are ten topologically
unique diagrammatic representations, and these are shown in 
Fig. \ref{slater_fig}(a-j). The diagrams in the figure correspond to, following
the notations introduced earlier, $v_{ab}^{pq} $, $v_{ra}^{qp}$, 
$v_{ba}^{cp}$, $v_{ar}^{cp}$, $v_{ra}^{cp}$, $v_{ab}^{cd}$, $v_{ar}^{cd}$, 
$v_{rs}^{pq}$, $v_{rs}^{pc}$  and $v_{rs}^{cd}$. While evaluating the terms
in the cluster equations, the lines below the interaction line (dashed lines)
or the orbitals with the indexes $r$, $s$, $c$ and $d$, contract
with the cluster operators.  We refer to these lines as the {\em internal} 
indices. The others, namely $p$, $q$, $a$ and $b$ are the {\em external} 
indices. An advantage of these classifications  and definitions is, one 
immediately notices that the Slater integrals with more than two {\em external} 
lines do not contribute to the $T_1$ equation. In particular, the Slater 
integrals which do not contribute to $T_1$ are $v_{ab}^{pq} $, 
$v_{ra}^{qp}$ and $v_{bc}^{cp}$ but, all the Slater integrals contribute to 
the $T_2$ equation.

   In the second tier, the loops are grouped depending on the Slater 
integrals. For example, in the linearized RCC, $H_{\rm N}T_1^{(0)}$ 
contributes through two channels of contractions $v_{rb}^{pq}t_a^r$ and 
$v_{ab}^{cq}t_c^p$. Here, as mentioned earlier, the product of the operators 
$H_{\rm N}T_1^{(0)}$ imply all possible contractions. The two terms arise from 
two different types of Slater integrals. A more complicated example is the 
contribution from $v_{rs}^{pc}$, it does not contribute to the $T_2$ equations 
at the linear level but contributes through the nonlinear terms 
$H_{\rm N}T_2^{(0)}T_1^{(0)}$ and $H_{\rm N}T_1^{(0)}T_1^{(0)}T_1^{(0)}$. 
This grouping of diagrams based on the Slater integrals is equivalent of
{\em integral} driven in a limited sense. Collectively, we refer to the 
two-tiered loop structure as the {\em cluster-integral} driven.


\subsection{Memory parallel integral storage}

  The RCC equations are nonlinear algebraic equations and are solved 
iteratively using standard numerical methods. In the present work, we use
Jacobi iteration and convergence is accelerated with DIIS \cite{pulay-80}. 
For improved performance, we store the Slater integrals in memory (RAM). 
The storage of the four particle integrals $v_{rs}^{pq}$, however, require 
very large memory. For example, in the present calculations the number of the 
particle states $N_v > 100$ and the order of memory required to store all 
the $v_{rs}^{pq}$ in double precision scales as $O[10N_v^4]= O[10^9]$ bytes. 
Where, the factor of ten accounts for the eight bytes to represent a 
real number in double precision, and the number of multipoles in each 
integral. This is a conservative estimate, the actual requirement may exceed
this by a factor between five and ten.  Although the memory required is 
manageable with current technologies, it is still a large requirement.

  With a straight forward and trivial parallelization, it is possible to 
divide and parallelize the compute intensive part of the calculations. In a 
distributed memory environment or cluster computers, one of the more prevalent
architecture within the high performance computing community, the storage
of $v_{rs}^{pq}$ has a large memory foot print. More over, the same set
of integrals are stored across all nodes and leads to replication of data. 
This is rather expensive and could be a severe bottle neck to exploit parallel
computing for calculations with large basis sizes. In the present work, we 
present an implementation where there are no memory replications while storing 
$v_{rs}^{pq}$. In other words, the storage of the $v_{rs}^{pq}$ is distributed
across the nodes or done in parallel. We refer to this scheme as the 
{\em memory parallel } implementation. 

  The {\em memory parallel } storage of the $v_{rs}^{pq}$ takes advantage
of the {\em cluster-integral} driven structure of the code. The implementation
exploits one feature of this structure: the orbitals lines
of the {\em external} loops are not contracted. So, we can 
parallelize any of the external loops, as it is common to all the cluster 
diagrams. However, for improved performance we choose the particle line for 
parallelization. This ensures nearly equal workload for all the nodes as the 
number of particle states is usually an order magnitude larger than the number 
of core states. For the present discussion, let us assume that the $p$ loop is 
parallelzed across $N_n$ processors of a distributed memory system. In general,
for the type of calculations we are interested $N_n < N_v$, and each
of the processors, then, store $O[(10N_v^4)/N_n]= O[10^9/N_n]$ of the
four particle Slater integrals $v_{rs}^{pq}$. For example, if 
$5N_n\approx N_v$, a condition met in most of our routine computations, 
we get the memory required per processor as $\approx 2\times 10^8 $. This is
less than one gigabyte and hence, we can increase the basis set size without
memory constraints.
%
%
\begin{figure}[h]
  \centering
  \includegraphics[width = 9.0 cm]{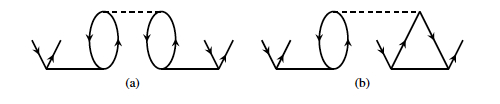}                                 
  \caption{The example diagrams which contribute to the RCC term
           $H_{\rm N}T_2T_2$ for doubles. Diagram (b) is the exchange at the
           vertex of $T_2$.}                   
 \label{ims_cc_fig}
\end{figure}
%
%


\subsection{Intermediate storage}

The group of CC diagrams for $T_2$ which arise from the Slater integral
$v^{pq}_{ab}$ involves four contractions, total of eight orbital lines and
summation over three multipoles. The evaluation of these diagrams requires
the maximum number of loops, and hence the CPU time. Consider, for example, 
one of the nonlinear terms 
$ \langle \Phi^{pq}_{ab}|H_{\rm N}T_2T_2 |\Phi_0 \rangle$, which is 
quadratic in $T_2$. There are 22 diagrams which contribute to this term and 
two are shown in the Fig \ref{ims_cc_fig}. The second diagram in this
figure arise due to exchange at one of the $T_2$.
%
%
\begin{figure}[h]
  \centering
  \includegraphics[width = 9.0 cm]{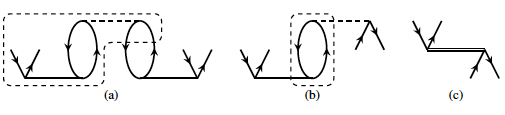}                                 
  \caption{The representation of 
          (a) a common part of the RCC diagram,
          (b) the IMS diagram and (c) the effective operator diagram.}
  \label{ims_diag_fig}
\end{figure}

   In both the diagrams, the number of hole and particle orbital lines 
are four each, and three multipole lines. The total number of operations 
(NOP) required to evaluate the diagram in Fig. \ref{ims_cc_fig}(a) is
${N_h}^4 {N_p}^4 {N_k}^3$. Here, $N_h$, $N_p$ and $N_k$ represent the
number of holes, particles and multipoles used in the computations, 
respectively. Considering the case of Na$^+$ as an
example, $N_h$ is 4, for a reasonable basis size $N_p$ can be
100, and since we include orbitals up to $h$-symmetry we may take
$N_k=11$. The total NOP is then $\approx 3.4 \times 10^{13}$. 
An important point to note is, this number corresponds to Na$^+$ which is a 
lighter ionic system. In the case of high-$Z$ atoms, where the number of holes 
states are large, the total NOP may increase significantly. 
%
%
\begin{figure}[h]
  \centering
  \includegraphics[width = 9.0 cm]{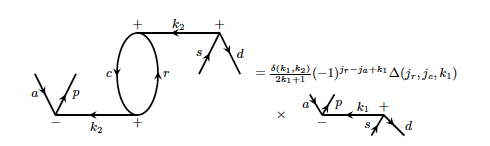}                                 
  \caption{The angular reduction of the IMS diagram. The free part on the
           right-hand side indicates the form of the effective operator.}
  \label{ims_angfac_fig}
\end{figure}

\begin{figure}[h]
  \centering
  \includegraphics[width = 9.0 cm]{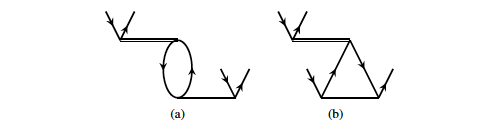}                                 
  \caption{The diagrams arise when effective operator is contracted with 
           $T_2$. (a) Equivalent to Fig. \ref{ims_cc_fig}(a).
           (b) Equivalent to the exchange diagram Fig. \ref{ims_cc_fig}(b).}
  \label{ims_eff_fig}
\end{figure}

     One way to reduce the computational time is through the use of IMS
scheme. In this approach, a common part of the CC diagrams is identified and
calculated separately. This is then stored as an effective operator. This
effective operator latter contracts with the CC operator to provide the 
contribution equivalent to the actual CC diagrams. The common part is referred 
to as the intermediate storage (IMS) diagram.
As shown in the Fig. \ref{ims_diag_fig}(a), the portion within the
dashed line is common to both diagrams in Fig. \ref{ims_cc_fig}, and 
is therefore an IMS diagram. The common portion is shown as a separate diagram
in Fig. \ref{ims_diag_fig}(b). The next step is to evaluate this IMS diagram 
and store in the form of an effective operator. The angular reduction of the 
IMS diagram is shown in the Fig. \ref{ims_angfac_fig}, where the removal 
of the closed loop has reduced it to angular factors and a diagram of free
lines. It is to be observed that the diagram with free lines is topologically 
similar to the effective operator shown in Fig. \ref{ims_diag_fig}(c). The 
final step to evaluate the CC diagram is to contract the effective operator 
with the $T_2$ operator. The diagrams which arise from the contractions are 
shown in the Fig. \ref{ims_eff_fig}. These are equivalent to the
diagrams in the Fig. \ref{ims_cc_fig}.
%
%

    After the implementation of computations with the IMS diagrams, the total 
NOP required to compute the diagram in Fig. \ref{ims_cc_fig}(a) is the sum of 
NOP to compute the diagrams in Fig. \ref{ims_diag_fig}(b) and 
Fig. \ref{ims_eff_fig}(a). Using the analysis discussed earlier, this is
$2 N^3_h N^3_p N^2_k$, which smaller by a factor of $1/2N_hN_pN_k$. For the 
example of Na$^+$, the NOP required is $\approx 1.5 \times 10^{10}$.
The advantage of using IMS becomes more evident when we consider both
diagrams in the Fig. \ref{ims_cc_fig}. The total NOP require is 
$3 N^3_h N^3_p N^2_k$, which for Na$^+$ example is 
$\approx 2.3 \times 10^{10}$. However, without the use of IMS scheme it
is $\approx 6.8 \times 10^{13}$.


\section{Description of RCCPAC}

\subsection{Input data files}
 The package requires two input data files: the orbital basis file; and
the data file which provides information about the type of orbitals. 

{\em Orbital file}--The default name of the orbital file is \texttt{wfn.dat} 
and it is in binary format. The contents of the file are accessed with a call 
to the subroutine in \texttt{readorb.f}. It has the following data: \\

\noindent\underline{First two records}\\
The first two records pertains to the spatial
grid on which the orbitals are defined. The first record has the grid 
parameters {\tt h} and {\tt n}. The second record contains the 
arrays {\tt r}, {\tt rp} and {\tt rpor}. For the exponential grid used, the 
first is the grid point, second is the scaling factor required in the 
integration and last is the {\tt rp/r}. These two records are accessed by the
subroutine in \texttt{readorb.f} as follows:
\begin{verbatim}
  read(WFNIN)h, n
  read(WFNIN)(r(i), i = 1, n), (rp(i), i = 1, n), (rpor(i), i = 1, n)
\end{verbatim}

\noindent\underline{Remaining records}\\
 In the remaining part of the file, there are 
two records for each orbital. The first is the orbital energy, and the second 
stores the arrays of the large and small components of the orbitals. The 
records are grouped into orbitals of the same symmetries with increasing $j$. 
For example, the core orbitals are accessed in {\tt readorb.f} as follows:
\begin{verbatim}
   read(WFNIN)eorb(indx1)
   read(WFNIN)(pf(ii,indx1),ii=1,n),(qf(ii,indx1),ii=1,n)
\end{verbatim}
and then, the virtual orbitals of the same symmetry are read next. This is 
repeated till orbitals of all the symmetries are read.

{\em Basis and option file}--The default name of the data file which has the 
information about the orbital basis is \texttt{rccpac.in}. It is an ASCII file 
and consists of the following lines:
atomic weight, number of symmetries, and total number of orbitals, valence and
core of each symmetry. For the closed-shell case, the number of valence 
orbitals is zero and the information is not really required. We, however,
introduce it, so that the code may be upgraded for one-valence systems with
minimal modifications. As an example, consider the case of atomic Na, the 
contents of \texttt{rccpac.in}  for computation with an orbital set
consisting of nine symmetries is as given below:
\begin{verbatim}
22.99
2
9
19   0    2
15   0    1
15   0    1
13   0    0
13   0    0
11   0    0
11   0    0
9    0    0
9    0    0
\end{verbatim}
The entry ( {\tt 22.99}) in the first line is the atomic weight of $^{23}$Na 
and the next line is the {\tt option} of the computation. The various possible
values of  {\tt option} are: 2 for the closed-shell cluster amplitude 
computations; 4 for the one-valence cluster amplitude computations; and 8 for 
the one-valence properties computations. To combine the computations, the 
{\tt option} of the individual cases must be summed. For example, the value of
{\tt option} to compute the cluster amplitudes of the closed-shell and 
one-valence sector is 6. The sum of 2 and 4, the values of {\tt option} to do 
the computations of closed-shell and one-valence cluster amplitudes. 
The next line gives the number of symmetries (9) considered in the present 
computation. The symmetries are namely, $s_{1/2}$, $p_{1/2}$, $p_{3/2}$,
$d_{3/2}$, $d_{5/2}$, $f_{5/2}$, $f_{7/2}$, $g_{7/2}$ and $g_{9/2}$. The
third line provides information about the $s_{1/2}$ orbitals in the basis.
In this line, the entry {\tt 19} is the total number of orbitals in $s_{1/2}$ 
symmetry, and the other two entries {\tt 0} and {\tt 2} are the number of 
valence and core orbitals in the $s_{1/2}$ symmetry. Similarly, the remaining 
lines provide information about the orbitals in the remaining symmetries and 
in the sequence listed earlier. 

Next, as an example, we provide the contents of the input file of one-valence 
computations for $^{133}$Cs. In this case the value of {\tt option} is set as
14, which is the sum of 2, 4, and 8. These are the options corresponding to the 
computations of close-shell and one-valence cluster amplitude, and one-valence
properties. The contents of the input file is as given below:
\begin{verbatim}
132.91
14
9
17  1  5
13  1  4
13  1  4
13  1  2
13  1  2
11  0  0
11  0  0
11  0  0
11  0  0
13  0  0
13  0  0
13  0  0
13  0  0
\end{verbatim}
The key difference of the input file compared to the previous 
example is the inclusion of data about the valence shells, the non-zero
values in the second column on information about the basis functions. The 
non-zero values in the present case represent the valence orbitals of 
$6s$, $6p_{1/2}$, $6p_{3/2}$, $5d_{3/2}$, and $5d_{5/2}$, respectively.


\subsection{Constant parameters}
 The dimension of the arrays and various other parameters required in
different sections of the package are defined as parameters in the module 
{\tt param}. The module is part of the main subroutine file {\tt rccpac.f} 
and in the present version of the package the module is defined as follows:
\begin{verbatim}
      module param
        integer, parameter :: NHO = 27,  NPO = 170, MXL = 25, MXV = 20,
     &                        MDIM = 6000000, MN  = 950, MNSYM = 13,
     &                        MNS = 13, MXVR  = (MXV+1)/2, MNBAS = NHO+NPO, 
     &                        MNOCC = NHO, MNEXC =  NPO, 
        integer, parameter :: STDIN = 5, WFNIN = 7,NTFILE = 16,
     &                        STDOUT = 8,MASTER = 0, STDIMS = 9, NITMAX = 50, 
     &                        NPMAX = 128, PUNCH = 17

        real (8), parameter:: SMALL = 1.2d-8
      end module param
\end{verbatim}
In the module, the first set of parameters define the maximum number of  
a data set and these are as follows:\\
\hspace*{0.5cm}
\begin{tabular} {ll}
  \texttt{NHO}:   & core orbitals, \\
  \texttt{NPO}:   & virtual orbitals,\\
  \texttt{MXL}:   & multipoles of the cluster amplitudes,\\
  \texttt{MXV}:   & multipoles of the two-electron interaction,\\
  \texttt{MDIM}:  & cluster amplitudes, \\
  \texttt{MN}:    & grid points used to define the orbitals, \\
  \texttt{MNSYM}: & symmetries of orbitals,\\
  \texttt{MNS}:   & symmetries of orbitals, \\
\end{tabular}
\newline
The second group of parameters are related to I/O and iterations 
used in the computations. These are: \\
\hspace*{0.5cm}
\begin{tabular} {ll}
  \texttt{STDIN}:   & Unit number of the input file,\\
  \texttt{WFNIN}:   & Unit number of the orbital data file,\\
  \texttt{NTFILE}:  & \\
  \texttt{STDOUT}:  & Unit number of default output,\\
  \texttt{MASTER}:  & identity of the master in the MPI execution of 
                      the package,\\
  \texttt{STDIMS}:  &  \\
  \texttt{NITMAX}:  & maximum number of iteration in the Jacobi method to 
                      solve the cluster equations, \\
  \texttt{NPMAX}:   & \\
\end{tabular}
\newline The last element in the module, \texttt{SMALL}, is the parameter 
which is used to define the convergence criterion.


\subsection{Output data}

 On the successful completion the package generates the coupled-cluster
amplitudes and these are stored in the binary data file 
\texttt{ccamp\_0v.dat}. The information related to computation are given in 
the output file \texttt{rccpac.out}. The file has data about the orbital 
basis, number of two-electron integrals, number of IMS diagrams in each 
group, convergence parameter, and details of the DIIS computation.

\begin{verbatim}
    *****************************************************************
    *****************************************************************
                RELATIVISTIC COUPLED-CLUSTER PROGRAM
                               for
                       ATOMIC CALCULATIONS
                             (RCCPAC)
        Relativistic coupled-cluster theory with single and double
        excitation approximation is implemented in this code.    
        The cluster equations are solved using Jacobi iteration  
        with Direct Inversion in the Iterative Subspace (DIIS) to
        accelerate the convergence.                              

        Written by                                               

        Brajesh K. Mani              Physical Research Laboratory
        Siddhartha Chattopadhyay     Theoretical Physics Division
        Dilip Angom                  Navarangpura, Ahmedabad--09 
                                     Gujarat, INDIA              

    *****************************************************************
    *****************************************************************

DATE :
 Tue Jun 21 13:27:27 2016
 The number of orbitals in the core (ncore) =   4
                             valence (nval) =   0
                            occupied (nocc) =   4
                           virtual (nexcit) = 111
                             total (nbasis) = 115

-------------------------------------------------------------------------------
               ++Completed RCC (T0) skip calculations (symm.f)++
-------------------------------------------------------------------------------
 Number of single excited cluster amplitudes:         62
           double excited cluster amplitudes:      27364

-------------------------------------------------------------------------------
               ++Completed reading radial wave functions++ 
-------------------------------------------------------------------------------
  Core orbitals
  -------------
  Seq no.             Orbital Energy

     1                  -40.82654629
     2                   -3.08240070
     3                   -1.80141924
     4                   -1.79401059

  ----------------
  Virtual orbitals
  ----------------
  Seq no.             Orbital Energy

     5                   -0.18203250
     6                   -0.07016031
     7                   -0.03703966
     8                   -0.01737279
     9                    0.05584768
    10                    0.31893785
   ***                  ************
   113                   14.53407907
   114                   38.87397690
   115                  105.24442393
-------------------------------------------------------------------------------
        ++Entering coul_tab to tabulate two-electron Coulomb integrals++
-------------------------------------------------------------------------------

 The maximum number of <ph|v|hp> ( nskip_phhp) is:        54204
                       <hh|v|hh> ( nskip_hhhh) is:          109
                       <ph|v|ph> ( nskip_phph) is:        50252
                       <pp|v|pp> ( nskip_pppp) is:     78252104
                       <hh|v|pp> ( nskip_hhpp) is:        54204
                       <pp|v|hp> ( nskip_pphp) is:      1945909
                       <hp|v|pp> ( nskip_hppp) is:      1945909
                       <ph|v|hh> ( nskip_phhh) is:         1785
                       <hh|v|ph> ( nskip_hhph) is:         1785

 init_close1, nsing           0          62
-------------------------------------------------------------------------------
             ++Linearised unperturbed closed-shell (ldrivert0_0v)++
-------------------------------------------------------------------------------
 Convergence parameter is   0.120000D-07

 Iteration   1
       conv and convd are =   0.615626D-01      0.138845D+01
       eps  and epsd  are =   0.992945D-03      0.507402D-04

 Iteration   2
       conv and convd are =   0.415324D-02      0.270568D+00
       eps  and epsd  are =   0.669878D-04      0.988775D-05

       DIIS matrix elements
           1    1  0.139833D-02
           1    2 -0.139262D-04
           2    2  0.393844D-04
       DIIS solution
             0.363755D-01      0.963625D+00      0.374452D-04

 *********   ************      ***********      *************

 Iteration   7
       conv and convd are =   0.211860D-04      0.157455D-03
       eps  and epsd  are =   0.341710D-06      0.575409D-08

 Converged in **  7** iterations


 Maximum number of nskip_ims4p is:    156081572
                   nskip_ims4h is:          406
                 nskip_ims2p2h is:       212722
                 nskip_imsphhh is:         7293
                 nskip_imspphp is:      7727007

-------------------------------------------------------------------------------
             ++Nonlinear unperturbed closed-shell (nldrivert0_0v)++
-------------------------------------------------------------------------------
 Convergence parameter is   0.120000D-07

 Iteration   1
       conv and convd are =   0.204806D-02      0.125925D+00
       eps  and epsd  are =   0.330333D-04      0.460186D-05

 Iteration   2
       conv and convd are =   0.628172D-03      0.219799D-01
       eps  and epsd  are =   0.101318D-04      0.803242D-06

       DIIS matrix elements
           1    1  0.149170D-04
           1    2  0.146607D-05
           2    2  0.390725D-06
       DIIS solution
            -0.868926D-01      0.108689D+01      0.297286D-06
 
*********   ************      ***********      *************

 Iteration   5
       conv and convd are =   0.165985D-04      0.301626D-03
       eps  and epsd  are =   0.267719D-06      0.110227D-07

       DIIS matrix elements
           1    1  0.112876D-08
           1    2  0.302626D-09
           2    2  0.102218D-09
       DIIS solution
            -0.320281D+00      0.132028D+01      0.380308D-10
 Converged in **  5** iterations


-------------------------------------------------------------------------------
                        ++Compute correlation energy++
-------------------------------------------------------------------------------

 Correlation energy ( in atomic units):
       Contribution from direct   diagrams:   -0.5781920
       Contribution from exchange diagrams:    0.2088729
                                     Total:   -0.3693191
 
DATE :
 Tue Jun 21 13:41:38 2016
\end{verbatim}
In the printout of the output file \texttt{rccpac.out}, the rows 
with \texttt{***} indicate additional lines of data, but not included in 
the above for compactness. Albeit, we have given the correlation energy
as an example of the property computed using the CCSD wavefunction, any
other properties of a closed-shell atom or ion may be computed using the
cluster amplitudes in the output file \texttt{ccamp\_0v.dat}. 

For the case of one-valence systems, the contents of the output file of the 
properties computations of $^{133}$Cs is given below as an example. The 
computation of the E1 reduced matrix elements is given as an example of the 
one-valence properties. The hyperfine structure constants can also be 
computed in the same way.
\begin{verbatim}


    *****************************************************************
    *****************************************************************
                RELATIVISTIC COUPLED-CLUSTER PROGRAM
                               for
                       ATOMIC CALCULATIONS
                             (RCCPAC)


        Relativistic coupled-cluster theory with single and double
        excitation approximation is implemented in this code.    
        The cluster equations are solved using Jacobi iteration  
        with Direct Inversion in the Iterative Subspace (DIIS) to
        accelerate the convergence.                              


        Written by                                               

        Brajesh K. Mani              Physical Research Laboratory
        Siddhartha Chattopadhyay     Theoretical Physics Division
        Dilip Angom                  Navarangpura, Ahmedabad--09 
                                     Gujarat, INDIA              

    *****************************************************************
    *****************************************************************



DATE :
 Mon Oct 17 12:24:45 2016
 The number of orbitals in the core (ncore) =  17
                             valence (nval) =   5
                            occupied (nocc) =  22
                           virtual (nexcit) =  96
                             total (nbasis) = 113

-------------------------------------------------------------------------------
               ++Completed RCC (T0) skip calculations (symm.f)++
-------------------------------------------------------------------------------
 Number of single excited cluster amplitudes:        228
           double excited cluster amplitudes:     868773


-------------------------------------------------------------------------------
               ++Completed reading radial wave functions++ 
-------------------------------------------------------------------------------
  Core orbitals
  -------------
  Seq no.             Orbital Energy

     1                -1330.11726503
     2                 -212.56429987
     3                  -45.96972982
     4                   -9.51278742
     5                   -1.48980491
     6                 -199.42944490
     7                  -40.44831432
     8                   -7.44627286
     9                   -0.90789762
    10                 -186.43662752
    11                  -37.89433600
    12                   -6.92099140
    13                   -0.84033940
    14                  -28.30956867
    15                   -3.48562841
    16                  -27.77522550
    17                   -3.39691055

  ----------------
  Valence orbitals
  ----------------
  Seq no.             Orbital Energy

    18                   -0.12736841
    19                   -0.08561350
    20                   -0.08378191
    21                   -0.06440672
    22                   -0.06451711

  ----------------
  Virtual orbitals
  ----------------
  Seq no.             Orbital Energy

    23                   -0.05283804
    24                    0.02231458
    25                    0.24902056
    26                    0.92053859
    27                    2.87599696
   ***                  ************
   111                   70.38575916
   112                  207.56873070
   113                  611.39136250
-------------------------------------------------------------------------------
        ++Entering coul_tab to tabulate two-electron Coulomb integrals++
-------------------------------------------------------------------------------


 The maximum number of <ph|v|hp> ( nskip_phhp) is:      1734444
                       <hh|v|hh> ( nskip_hhhh) is:        97234
                       <ph|v|ph> ( nskip_phph) is:      1697124
                       <pp|v|pp> ( nskip_pppp) is:     53166820
                       <hh|v|pp> ( nskip_hhpp) is:      1734444
                       <pp|v|hp> ( nskip_pphp) is:      9290442
                       <hp|v|pp> ( nskip_hppp) is:      9290442
                       <ph|v|hh> ( nskip_phhh) is:       353553
                       <hh|v|ph> ( nskip_hhph) is:       353553

 init_close1, nsing           0         228
********************************************************************************
                     Output from the closed-shell part
********************************************************************************
-------------------------------------------------------------------------------
             ++Linearised unperturbed one-valence (ldrivert0_1v)++
-------------------------------------------------------------------------------
 Convergence parameter is   0.120000D-07
       Attachement energies
       Valence Orb       Corr Energy       Orb Energy       Attach Energy
            18          -0.166919D-01     -0.127368D+00    -0.144060D+00
            19          -0.655392D-02     -0.856135D-01    -0.921674D-01
            20          -0.592481D-02     -0.837819D-01    -0.897067D-01
            21          -0.100909D-01     -0.644067D-01    -0.744976D-01
            22          -0.972703D-02     -0.645171D-01    -0.742441D-01

 Iteration   1
       conv and convd are =   0.229040D+01      0.630055D+02
       eps  and epsd  are =   0.100456D-01      0.725224D-04

       Attachement energies
       Valence Orb       Corr Energy       Orb Energy       Attach Energy
            18          -0.134525D-01     -0.127368D+00    -0.140821D+00
            19          -0.544351D-02     -0.856135D-01    -0.910570D-01
            20          -0.492562D-02     -0.837819D-01    -0.887075D-01
            21          -0.763087D-02     -0.644067D-01    -0.720376D-01
            22          -0.738749D-02     -0.645171D-01    -0.719046D-01


 *********   ************      ***********      *************

 Iteration  10
       conv and convd are =   0.580033D-03      0.117220D-01
       eps  and epsd  are =   0.254401D-05      0.134926D-07

       Attachement energies
       Valence Orb       Corr Energy       Orb Energy       Attach Energy
            18          -0.169498D-01     -0.127368D+00    -0.144318D+00
            19          -0.692914D-02     -0.856135D-01    -0.925426D-01
            20          -0.620691D-02     -0.837819D-01    -0.899888D-01
            21          -0.136117D-01     -0.644067D-01    -0.780185D-01
            22          -0.128607D-01     -0.645171D-01    -0.773778D-01

 Iteration  11
       conv and convd are =   0.333555D-03      0.447109D-02
       eps  and epsd  are =   0.146296D-05      0.514645D-08

 Converged in ** 11** iterations


-------------------------------------------------------------------------------
             ++Nonlinear unperturbed one-valence (nldrivert0_1v)++
-------------------------------------------------------------------------------
 Convergence parameter is   0.120000D-07
       Attachement energies
       Valence Orb       Corr Energy       Orb Energy       Attach Energy
            18          -0.169508D-01     -0.127368D+00    -0.144319D+00
            19          -0.692949D-02     -0.856135D-01    -0.925430D-01
            20          -0.620722D-02     -0.837819D-01    -0.899891D-01
            21          -0.136117D-01     -0.644067D-01    -0.780184D-01
            22          -0.128606D-01     -0.645171D-01    -0.773777D-01

 Iteration   1
       conv and convd are =   0.410651D+00      0.359759D+01
       eps  and epsd  are =   0.180110D-02      0.414100D-05

       Attachement energies
       Valence Orb       Corr Energy       Orb Energy       Attach Energy
            18          -0.164705D-01     -0.127368D+00    -0.143839D+00
            19          -0.666249D-02     -0.856135D-01    -0.922760D-01
            20          -0.588440D-02     -0.837819D-01    -0.896663D-01
            21          -0.131382D-01     -0.644067D-01    -0.775449D-01
            22          -0.124401D-01     -0.645171D-01    -0.769572D-01

 *********   ************      ***********      *************

 Iteration   9
       conv and convd are =   0.492117D-03      0.474424D-02
       eps  and epsd  are =   0.215841D-05      0.546085D-08

       DIIS matrix elements
           1    1  0.585631D-06
           1    2  0.166541D-06
           1    3  0.146858D-06
           1    4  0.565027D-07
           2    2  0.259704D-06
           2    3  0.480579D-07
           2    4  0.887415D-07
           3    3  0.398448D-07
           3    4  0.155797D-07
           4    4  0.316634D-07
       DIIS solution
            -0.131591D+00     -0.321214D+00      0.549912D+00
             0.902893D+00      0.121586D-08
 Converged in **  9** iterations


       Attachement energies
       Valence Orb       Corr Energy       Orb Energy       Attach Energy
            18          -0.163371D-01     -0.127368D+00    -0.143705D+00
            19          -0.660373D-02     -0.856135D-01    -0.922172D-01
            20          -0.583622D-02     -0.837819D-01    -0.896181D-01
            21          -0.119459D-01     -0.644067D-01    -0.763527D-01
            22          -0.113554D-01     -0.645171D-01    -0.758725D-01
-------------------------------------------------------------------------------
               ++One-valence properties computations (prop_1v)++
-------------------------------------------------------------------------------
 Normalization

    State     Normalization factor

      18         0.993542
      19         0.996916
      20         0.997362
      21         0.993279
      22         0.993639

  Final state      Intial state                      E1 amplitude
       1                  2       DF value             -5.228273
                                  Total value          -4.494237

       1                  3       DF value             -7.360728
                                  Total value          -6.350515

       2                  1       DF value             -5.228273
                                  Total value          -4.500968

       2                  4       DF value              8.874962
                                  Total value           7.313672

       2                  5       DF value              0.000000
                                  Total value           0.000000

       3                  1       DF value              7.360728
                                  Total value           6.368736

       3                  4       DF value              4.015906
                                  Total value           3.293150

       3                  5       DF value             12.051371
                                  Total value           9.995387

       4                  2       DF value             -8.874962
                                  Total value          -7.327270

       4                  3       DF value              4.015906
                                  Total value           3.294046

       5                  2       DF value              0.000000
                                  Total value           0.000000

       5                  3       DF value            -12.051371
                                  Total value          -9.982575

 
DATE :
 Mon Oct 17 19:10:23 2016

\end{verbatim}
In the output file shown above, the states identified as 1, 2, 3, 4, and 5
correspond to the $6s(^2S_{1/2})$, $6p (^2P_{1/2})$, $6p(^2P_{3/2})$,
$5d(^2D_{3/2})$ and $5d(^2D_{5/2})$ states of Cs. The hyperfine structure
constants can also be computed in the same way by choosing the hyperfine
structure constant subroutine in the main driver subroutine \texttt{rccpac.f}
where the properties driver subroutine \texttt{prop\_1v.f} is called.
In the case of hyperfine structure constants, the output from the code gives
the matrix elements of the hyperfine interaction Hamiltonian in the electronic
sector. So, to obtain the hyperfine structure constants in the units
of MHz, the results obtain from the code should be multiplied by the 
gyromagnetic ratio of the atom or ion, and factor of 13074.69. For 
compactness we have not shown the contents of the output file for the hyperfine
structure constant computations.


\section*{Acknowledgments}

   We acknowledge the valuable discussions with S. A. Silotri, S. Gautam, 
and A. Roy on various topics related to many-body physics. We are
grateful to B. P. Das and D. Mukherjee for many discussions on the theoretical
details of coupled-cluster theory. We thank, Per Jonsson, Farid Parpia,
K. V. P. Latha, and B. P. Das for allowing us the use of subroutines they
have written as part of other scientific package. The example results shown
in the paper are based on the computations using the HPC cluster Vikram-100
at Physical Research Laboratory, Ahmedabad.

\section*{References}

\begin{thebibliography}{31}
\expandafter\ifx\csname natexlab\endcsname\relax\def\natexlab#1{#1}\fi
\providecommand{\bibinfo}[2]{#2}
\ifx\xfnm\relax \def\xfnm[#1]{\unskip,\space#1}\fi
\bibitem[{Coester(1958)}]{coester-58}
\bibinfo{author}{F.~Coester}, \bibinfo{journal}{Nucl. Phys.}
  \bibinfo{volume}{7} (\bibinfo{year}{1958}) \bibinfo{pages}{421 -- 424}.
\bibitem[{Coester and K{\"{u}}mmel(1960)}]{coester-60}
\bibinfo{author}{F.~Coester}, \bibinfo{author}{H.~K{\"{u}}mmel},
  \bibinfo{journal}{Nucl. Phys.} \bibinfo{volume}{17} (\bibinfo{year}{1960})
  \bibinfo{pages}{477 -- 485}.
\bibitem[{\v{C}\'{i}\v{z}ek(1966)}]{cizek-66}
\bibinfo{author}{J.~\v{C}\'{i}\v{z}ek}, \bibinfo{journal}{J. Chem. Phys.}
  \bibinfo{volume}{45} (\bibinfo{year}{1966}) \bibinfo{pages}{4256}.
\bibitem[{\v{C}\'{i}\v{z}ek(1969)}]{cizek-69}
\bibinfo{author}{J.~\v{C}\'{i}\v{z}ek}, \bibinfo{journal}{Adv. Chem. Phys.}
  \bibinfo{volume}{14} (\bibinfo{year}{1969}) \bibinfo{pages}{35--89}.
\bibitem[{Hagen et~al.(2014)Hagen, Papenbrock, Hjorth-Jensen, and
  Dean}]{hagen-14}
\bibinfo{author}{G.~Hagen}, \bibinfo{author}{T.~Papenbrock},
  \bibinfo{author}{M.~Hjorth-Jensen}, \bibinfo{author}{D.~J. Dean},
  \bibinfo{journal}{Rep. Prog. Phys.} \bibinfo{volume}{77}
  (\bibinfo{year}{2014}) \bibinfo{pages}{096302}.
\bibitem[{Gopakumar et~al.(2001)Gopakumar, Merlitz, Majumder, Chaudhuri, Das,
  Mahapatra, and Mukherjee}]{geetha-01}
\bibinfo{author}{G.~Gopakumar}, \bibinfo{author}{H.~Merlitz},
  \bibinfo{author}{S.~Majumder}, \bibinfo{author}{R.~K. Chaudhuri},
  \bibinfo{author}{B.~P. Das}, \bibinfo{author}{U.~S. Mahapatra},
  \bibinfo{author}{D.~Mukherjee}, \bibinfo{journal}{Phys. Rev. A}
  \bibinfo{volume}{64} (\bibinfo{year}{2001}) \bibinfo{pages}{032502}.
\bibitem[{Pal et~al.(2007)Pal, Safronova, Johnson, Derevianko, and
  Porsev}]{pal-07}
\bibinfo{author}{R.~Pal}, \bibinfo{author}{M.~S. Safronova},
  \bibinfo{author}{W.~R. Johnson}, \bibinfo{author}{A.~Derevianko},
  \bibinfo{author}{S.~G. Porsev}, \bibinfo{journal}{Phys. Rev. A}
  \bibinfo{volume}{75} (\bibinfo{year}{2007}) \bibinfo{pages}{042515}.
\bibitem[{Mani et~al.(2009)Mani, Latha, and Angom}]{mani-09}
\bibinfo{author}{B.~K. Mani}, \bibinfo{author}{K.~V.~P. Latha},
  \bibinfo{author}{D.~Angom}, \bibinfo{journal}{Phys. Rev. A}
  \bibinfo{volume}{80} (\bibinfo{year}{2009}) \bibinfo{pages}{062505}.
\bibitem[{Isaev et~al.(2004)Isaev, Petrov, Mosyagin, Titov, Eliav, and
  Kaldor}]{isaev-04}
\bibinfo{author}{T.~A. Isaev}, \bibinfo{author}{A.~N. Petrov},
  \bibinfo{author}{N.~S. Mosyagin}, \bibinfo{author}{A.~V. Titov},
  \bibinfo{author}{E.~Eliav}, \bibinfo{author}{U.~Kaldor},
  \bibinfo{journal}{Phys. Rev. A} \bibinfo{volume}{69} (\bibinfo{year}{2004})
  \bibinfo{pages}{030501}.
\bibitem[{Li et~al.(2014)Li, Bishop, and Campbell}]{li-14}
\bibinfo{author}{P.~H.~Y. Li}, \bibinfo{author}{R.~F. Bishop},
  \bibinfo{author}{C.~E. Campbell}, \bibinfo{journal}{Phys. Rev. B}
  \bibinfo{volume}{89} (\bibinfo{year}{2014}) \bibinfo{pages}{220408}.
\bibitem[{{\v{C}}{\'a}rsk{\`y} et~al.(2010){\v{C}}{\'a}rsk{\`y}, Paldus, and
  Pittner}]{carsky-10}
\bibinfo{author}{P.~{\v{C}}{\'a}rsk{\`y}}, \bibinfo{author}{J.~Paldus},
  \bibinfo{author}{J.~Pittner}, \bibinfo{title}{Recent Progress in Coupled
  Cluster Methods: Theory and Applications}, Challenges and Advances in
  Computational Chemistry and Physics, \bibinfo{publisher}{Springer},
  \bibinfo{year}{2010}.
\bibitem[{Bartlett and Musia\l{}(2007)}]{bartlett-07}
\bibinfo{author}{R.~J. Bartlett}, \bibinfo{author}{M.~Musia\l{}},
  \bibinfo{journal}{Rev. Mod. Phys.} \bibinfo{volume}{79}
  (\bibinfo{year}{2007}) \bibinfo{pages}{291--352}.
\bibitem[{Purvis and Bartlett(1982)}]{purvis-82}
\bibinfo{author}{G.~D. Purvis}, \bibinfo{author}{R.~J. Bartlett},
  \bibinfo{journal}{J. Chem. Phys.} \bibinfo{volume}{76} (\bibinfo{year}{1982})
  \bibinfo{pages}{1910--1918}.
\bibitem[{Nataraj et~al.(2008)Nataraj, Sahoo, Das, and Mukherjee}]{nataraj-08}
\bibinfo{author}{H.~S. Nataraj}, \bibinfo{author}{B.~K. Sahoo},
  \bibinfo{author}{B.~P. Das}, \bibinfo{author}{D.~Mukherjee},
  \bibinfo{journal}{Phys. Rev. Lett.} \bibinfo{volume}{101}
  (\bibinfo{year}{2008}) \bibinfo{pages}{033002}.
\bibitem[{Latha et~al.(2009)Latha, Angom, Das, and Mukherjee}]{latha-09}
\bibinfo{author}{K.~V.~P. Latha}, \bibinfo{author}{D.~Angom},
  \bibinfo{author}{B.~P. Das}, \bibinfo{author}{D.~Mukherjee},
  \bibinfo{journal}{Phys. Rev. Lett.} \bibinfo{volume}{103}
  (\bibinfo{year}{2009}) \bibinfo{pages}{083001}.
\bibitem[{Wansbeek et~al.(2008)Wansbeek, Sahoo, Timmermans, Das, and
  Mukherjee}]{wansbeek-08}
\bibinfo{author}{L.~W. Wansbeek}, \bibinfo{author}{B.~K. Sahoo},
  \bibinfo{author}{R.~G.~E. Timmermans}, \bibinfo{author}{B.~P. Das},
  \bibinfo{author}{D.~Mukherjee}, \bibinfo{journal}{Phys. Rev. A}
  \bibinfo{volume}{78} (\bibinfo{year}{2008}) \bibinfo{pages}{012515}.
\bibitem[{Sahoo et~al.(2009)Sahoo, Das, and Mukherjee}]{sahoo-09}
\bibinfo{author}{B.~K. Sahoo}, \bibinfo{author}{B.~P. Das},
  \bibinfo{author}{D.~Mukherjee}, \bibinfo{journal}{Phys. Rev. A}
  \bibinfo{volume}{79} (\bibinfo{year}{2009}) \bibinfo{pages}{052511}.
\bibitem[{Chattopadhyay et~al.(2012{\natexlab{a}})Chattopadhyay, Mani, and
  Angom}]{chattopadhyay-12a}
\bibinfo{author}{S.~Chattopadhyay}, \bibinfo{author}{B.~K. Mani},
  \bibinfo{author}{D.~Angom}, \bibinfo{journal}{Phys. Rev. A}
  \bibinfo{volume}{86} (\bibinfo{year}{2012}{\natexlab{a}})
  \bibinfo{pages}{022522}.
\bibitem[{Chattopadhyay et~al.(2012{\natexlab{b}})Chattopadhyay, Mani, and
  Angom}]{chattopadhyay-12b}
\bibinfo{author}{S.~Chattopadhyay}, \bibinfo{author}{B.~K. Mani},
  \bibinfo{author}{D.~Angom}, \bibinfo{journal}{Phys. Rev. A}
  \bibinfo{volume}{86} (\bibinfo{year}{2012}{\natexlab{b}})
  \bibinfo{pages}{062508}.
\bibitem[{Chattopadhyay et~al.(2013{\natexlab{a}})Chattopadhyay, Mani, and
  Angom}]{chattopadhyay-13a}
\bibinfo{author}{S.~Chattopadhyay}, \bibinfo{author}{B.~K. Mani},
  \bibinfo{author}{D.~Angom}, \bibinfo{journal}{Phys. Rev. A}
  \bibinfo{volume}{87} (\bibinfo{year}{2013}{\natexlab{a}})
  \bibinfo{pages}{042520}.
\bibitem[{Chattopadhyay et~al.(2013{\natexlab{b}})Chattopadhyay, Mani, and
  Angom}]{chattopadhyay-13b}
\bibinfo{author}{S.~Chattopadhyay}, \bibinfo{author}{B.~K. Mani},
  \bibinfo{author}{D.~Angom}, \bibinfo{journal}{Phys. Rev. A}
  \bibinfo{volume}{87} (\bibinfo{year}{2013}{\natexlab{b}})
  \bibinfo{pages}{062504}.
\bibitem[{Chattopadhyay et~al.(2014)Chattopadhyay, Mani, and
  Angom}]{chattopadhyay-14}
\bibinfo{author}{S.~Chattopadhyay}, \bibinfo{author}{B.~K. Mani},
  \bibinfo{author}{D.~Angom}, \bibinfo{journal}{Phys. Rev. A}
  \bibinfo{volume}{89} (\bibinfo{year}{2014}) \bibinfo{pages}{022506}.
\bibitem[{Mani and Angom(2011)}]{mani-11}
\bibinfo{author}{B.~K. Mani}, \bibinfo{author}{D.~Angom},
  \bibinfo{journal}{Phys. Rev. A} \bibinfo{volume}{83} (\bibinfo{year}{2011})
  \bibinfo{pages}{012501}.
\bibitem[{Lindgren and Morrison(1986)}]{lindgren-86}
\bibinfo{author}{I.~Lindgren}, \bibinfo{author}{J.~Morrison},
  \bibinfo{title}{Atomic Many-Body Theory}, \bibinfo{publisher}{Springer},
  \bibinfo{address}{Berlin}, \bibinfo{year}{2nd Edition, 1986}.
\bibitem[{Shavitt and Bartlett(2009)}]{shavitt-09}
\bibinfo{author}{I.~Shavitt}, \bibinfo{author}{R.~Bartlett},
  \bibinfo{title}{Many-Body Methods in Chemistry and Physics},
  \bibinfo{publisher}{Cambridge University Press},
  \bibinfo{address}{Cambridge}, \bibinfo{year}{2009}.
\bibitem[{Mani and Angom(2010)}]{Mani-10}
\bibinfo{author}{B.~K. Mani}, \bibinfo{author}{D.~Angom},
  \bibinfo{journal}{Phys. Rev. A} \bibinfo{volume}{81} (\bibinfo{year}{2010})
  \bibinfo{pages}{042514}.
\bibitem[{Schwartz(1955)}]{schwartz-55}
\bibinfo{author}{C.~Schwartz}, \bibinfo{journal}{Phys. Rev.}
  \bibinfo{volume}{97} (\bibinfo{year}{1955}) \bibinfo{pages}{380--395}.
\bibitem[{J{\"{o}}nsson et~al.(2007)J{\"{o}}nsson, He, Froese~Fischer, and
  Grant}]{jonsson-07}
\bibinfo{author}{P.~J{\"{o}}nsson}, \bibinfo{author}{X.~He},
  \bibinfo{author}{C.~Froese~Fischer}, \bibinfo{author}{I.~P. Grant},
  \bibinfo{journal}{Comp. Phys. Comm.} \bibinfo{volume}{177}
  (\bibinfo{year}{2007}) \bibinfo{pages}{597 -- 622}.
\bibitem[{Mohanty and Clementi(1990)}]{mohanty-90}
\bibinfo{author}{A.~K. Mohanty}, \bibinfo{author}{E.~Clementi},
  \bibinfo{journal}{J. Chem. Phys.} \bibinfo{volume}{93} (\bibinfo{year}{1990})
  \bibinfo{pages}{1829--1833}.
\bibitem[{Chaudhuri et~al.(1999)Chaudhuri, Panda, and Das}]{chaudhuri-99}
\bibinfo{author}{R.~K. Chaudhuri}, \bibinfo{author}{P.~K. Panda},
  \bibinfo{author}{B.~P. Das}, \bibinfo{journal}{Phys. Rev. A}
  \bibinfo{volume}{59} (\bibinfo{year}{1999}) \bibinfo{pages}{1187--1196}.
\bibitem[{Pulay(1980)}]{pulay-80}
\bibinfo{author}{P.~Pulay}, \bibinfo{journal}{Chem. Phys. Lett.}
  \bibinfo{volume}{73} (\bibinfo{year}{1980}) \bibinfo{pages}{393 -- 398}.

\end{thebibliography}

\end{document}